\documentclass[11pt,twoside]{article}
\usepackage{asp2010}
\usepackage{graphicx}

\def\ltsim{\mathrel{\hbox{\rlap{\hbox{\lower4pt\hbox{$\sim$}}}\hbox{$<$}}}}
\newcommand{\bz}{\ensuremath{\langle B_z\rangle}}

\resetcounters

\begin{document}

\title{The MiMeS Survey of Magnetism in Massive Stars}
\author{Gregg A. Wade,$^1$ Jason H. Grunhut,$^{1,2}$ and The MiMeS Collaboration}
\affil{$^1$Department of Physics, Royal Military College of Canada, Kingston, Ontario, Canada}
\affil{$^2$Department of Physics, Engineering Physics and Astronomy, Queen's University, Kingston, Ontario, Canada}

\begin{abstract}
The Magnetism in Massive Stars (MiMeS) survey represents a high-precision systematic search for magnetic fields in hot, massive OB stars.
To date, MiMeS Large Programs (ESPaDOnS@CFHT, Narval@TBL, HARPSpol@ESO3.6) and associated PI programs (FORS@VLT) 
have yielded nearly 1200 circular spectropolarimetric observations of over 350 OB stars. Within this sample, 20 stars are detected as magnetic.
Follow-up observations of new detections reveals (i) a large diversity of magnetic properties, (ii) ubiquitous evidence for magnetic wind
confinement in optical spectra of all magnetic O stars, and (iii) the presence of strong, organized magnetic fields in all known Galactic Of?p
stars, and iv) a complete absence of magnetic fields in classical Be stars.
\end{abstract}

\section{Introduction}

Massive stars are those stars with initial masses above about 8 times that of the sun, destined to end their lives in catastrophic explosions in the form of supernovae. These represent the most massive and luminous stellar component of the Universe, and are the crucibles in which the lionÕs share of the chemical elements are forged. These rapidly-evolving stars drive the chemistry, structure and evolution of galaxies, dominating the ecology of the Universe - not only as supernovae, but also during their entire lifetimes - with far-reaching consequences.

Massive OB stars are unique laboratories for investigating the physics of stellar magnetism. Magnetic fields have clear influence on their rotation rates \citep[rotation periods of many detected magnetic OB stars are significantly longer than those of non-magnetic OB stars of similar spectral types; e.g.][]{2009MNRAS.392.1022U, 2007MNRAS.381..433H, 2010MNRAS.407.1423M}. Evolutionary models \citep[][]{2003A&A...411..543M, 2004A&A...422..225M} and, recently, observations (Briquet et al., MNRAS, submitted) of massive stars suggest that the internal rotation profile is strongly modified by the presence of a magnetic field, enforcing essentially solid-body rotation throughout the bulk of the outer radiative zone. Magnetic fields have clear and fundamental effects on the structure, dynamics and heating of the powerful radiative winds of OB stars \citep[e.g.][]{2002ApJ...576..413U, 2012MNRAS.tmpL.433S}. The lives of magnetic OB stars are therefore expected to differ significantly from those of their non-magnetic brethren.

The Magnetism in Massive Stars (MiMeS) Project represents a comprehensive, multidisciplinary strategy by an international team of experts to address the Òbig questionsÓ related to the complex and puzzling magnetism of massive stars. In 2008, the MiMeS Project was awarded ÒLarge ProgramÓ (LP) status by both Canada and France (PI G.A. Wade) at the Canada-France-Hawaii Telescope (CFHT), where the Project was allocated 640 hours of dedicated time with the ESPaDOnS spectropolarimeter from late 2008 through 2013. Since then the MiMeS consortium has been awarded LP status with the Narval spectropolarimeter at the Bernard Lyot Telescope in France (a total of 590 hours, PI C. Neiner) and HarpsPol at ESOÕs 3.6 m telescope at La Silla, Chile (a total of 30 nights, PI E. Alecian). In addition to these LPs, the MiMeS Project is supported by numerous PI programs from such observatories as the Anglo-Australian Telescope, the Chandra X-Ray Observatory, the Dominion Astrophysical Observatory, the HST, MOST, SMARTS, and the Very Large Telescope (VLT).

The structure of the MiMeS LP includes approximately 40\% of the allocated hours reserved for a $\sim40$-target ÒTargeted Component" (TC) which obtains time-resolved high-resolution spectropolarimetry of known magnetic massive stars in unpolarized light (Stokes $I$), circularly polarized light (Stokes $V$), and for some targets also linear polarized light (Stokes $Q$ and $U$ ), with the goal to constrain detailed models of the surface magnetic field structure and related physics of these stars. In addition, 60\% of the allocation hours are dedicated to the $\sim 400$-target ÒSurvey Component" (SC), with the aim to search for magnetic fields in stars with no prior detection of a magnetic field, to provide critical missing information about field incidence and statistical field properties from a much larger sample of massive stars. The SC focuses primarily on Galactic field stars consisting of Ònormal" field OB stars, cluster OB stars, emission line B-type and classical Be stars, pre-main sequence Herbig Be stars, and Wolf-Rayet stars. The majority of the SC targets were chosen such that they have apparent magnitudes $V< 9$ and projected rotational velocities $v\sin i<300$~km/s to ensure sensitivity to relatively weak large-scale fields.

\,

The MiMeS Project has four main science objectives:
\begin{itemize}
\item To identify and model the physical processes responsible for the generation of magnetic fields in massive stars;
\item To observe and model the detailed interaction between the magnetic fields and stellar winds of massive stars;
\item To investigate the role of the magnetic field in modifying the rotational properties of massive stars;
\item To investigate the impact of magnetic fields on massive star evolution, and the connection between magnetic fields of non-degenerate massive stars and their descendants, the neutron stars and magnetars.
\end{itemize}

\begin{figure}
\begin{centering}
\includegraphics[width=5.2in]{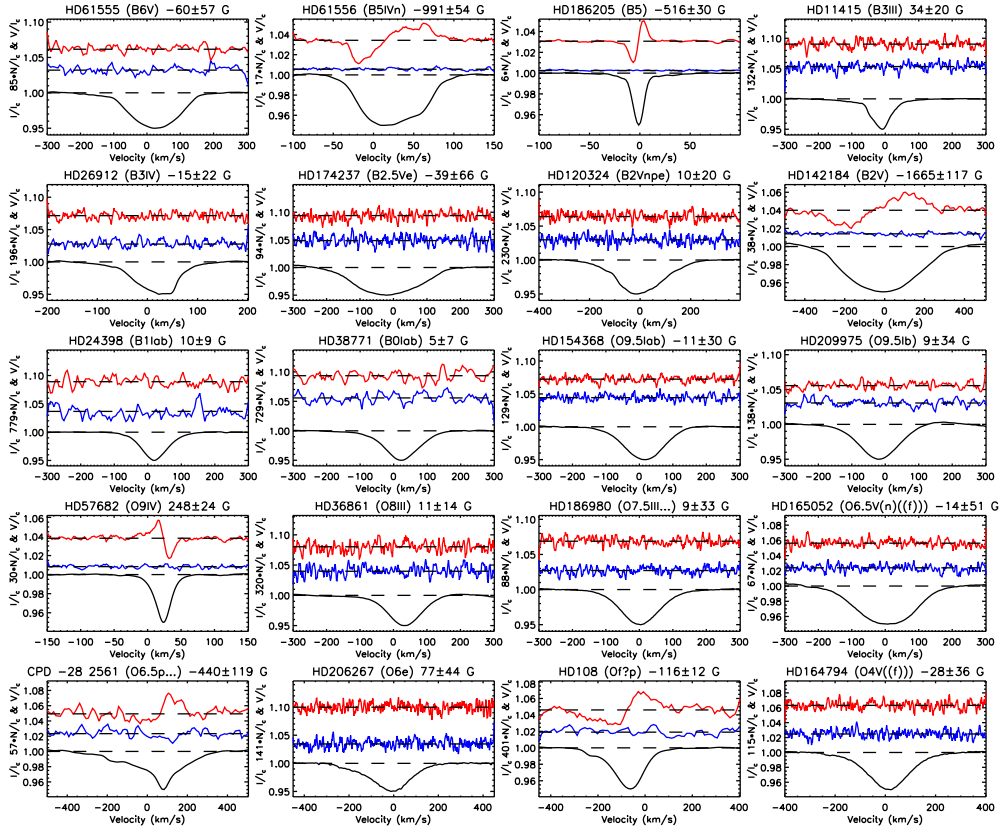}
\caption{Mean LSD circularly polarized Stokes V (top curve, red), diagnostic null (middle curve, blue) and unpolarized Stokes I (bottom curve, black) profiles for a small sample of the Survey Component observations. The spectral lines have all been scaled such that the unpolarized absorption lines have a depth that is 5\% of the continuum and the null and Stokes V profiles have been shifted and amplified by the indicated factor for display purposes. A magnetic field is detected if there is excess polarization signal detected within the line profile of the Stokes V spectrum. Clear magnetic Zeeman signatures are visible in HD 61556, HD 186205, HD 142184, HD 57682, CPD -28 2561 and HD 108 while no Zeeman signature is present in the other stars. Also included are the spectral type and computed longitudinal field for each LSD profile.}
\end{centering}
\end{figure}

\section{MiMeS SC observations}

The instrument employed for the ESPaDOnS SC is the bench-mounted, high throughput, fibre-fed, high-resolution ($R\simeq 65000$) ESPaDOnS spectropolarimeter. The spectrograph is both thermally and vibrationally isolated to reduce systematic effects, which could result in spurious magnetic signatures. The spectrograph is a cross-dispersed \'echelle spectrograph capable of obtaining near complete coverage of the optical spectrum (370 to 1050 nm) in a single exposure. The polarimeter is composed of one quarter-wave and two half-wave Fresnel rhombs with a Wollaston prism to provide achromatic polarization in all four Stokes parameters. Our exposure times are chosen to statistically provide signal-to-noise ratios (S/Ns) high enough to detect dipole magnetic fields with a range of target surface polar field intensities from 100 G to 2 kG, depending principally on the stellar spectral properties. The construction and operation of the Narval and HARPSpol instruments (employed respectively for the LPs at the TBL and ESO 3.6m telescope) are fundamentally similar to ESPaDOnS.

To diagnose the magnetic field, the SC relies on the magnetic splitting of photospheric spectral lines (the Zeeman effect). Since the magnetic splitting observed in unpolarized light is difficult to detect in hot stars due to their intrinsically broad spectral lines (from thermal, turbulent or rotational broadening), we rely on the Zeeman signatures that are produced in circularly polarized light (Stokes $V$), which can still be observed even in spectral lines with significant rotational or turbulent broadening. While the line splitting observed in the unpolarized light conveys information about the total modulus of the magnetic field, the Zeeman signature in circular polarization relays information about the line-of-sight (LOS) component of the magnetic field. It is from the Zeeman signature that we obtained a quantitative measure me of the disc-integrated LOS component, the so-called mean longitudinal magnetic field \bz. The relative strength of the Zeeman signature (compared to the continuum level) depends only on the magnetic field strength, the total broadening, and the atomic properties of the spectral line (line depth and magnetic sensitivity). Our ability to resolve these signatures across individual line profiles in metallic and helium lines is only possible because of the high spectral resolution of the ESPaDonS, Narval and HARPSpol polarimeters. This allows us to unambiguously detect the presence of magnetic signatures, but requires sufficiently high S/Ns in order to distinguish the signatures relative to the noise level. Furthermore, the ability to resolve individual line profiles allows us to (for most magnetic configurations) still detect a Zeeman signature even when the net LOS component of the magnetic field is consistent with zero. 

Since the detected magnetic fields in most hot, massive stars are relatively weak, we require S/Ns on the order of $\sim 10\,000$ to detect Zeeman signatures in individual optical spectral lines of typical O- or B-type stars. In order to increase the S/N and enhance our sensitivity to weak Zeeman signatures, we employ the Least-Squares Deconvolution (LSD) multi-line technique of \citet{1997MNRAS.291..658D} to produce a single mean profile from the unpolarized light, circularly polarized light and also from the diagnostic null spectrum (the null spectrum is computed in such a way that the real stellar polarization cancels out, and is used to diagnose the presence of spurious signatures resulting from instrumental or other systematic effects). The S/N increase is dependent on the number of spectral lines used in the analysis. LSD profiles are shown in Fig. 1 for a large sample of MiMeS SC stars with different spectral types, spectral characteristics and magnetic properties. 

\section{Quality control}

\begin{figure}
\begin{centering}
\includegraphics[width=2.2in,angle=-90]{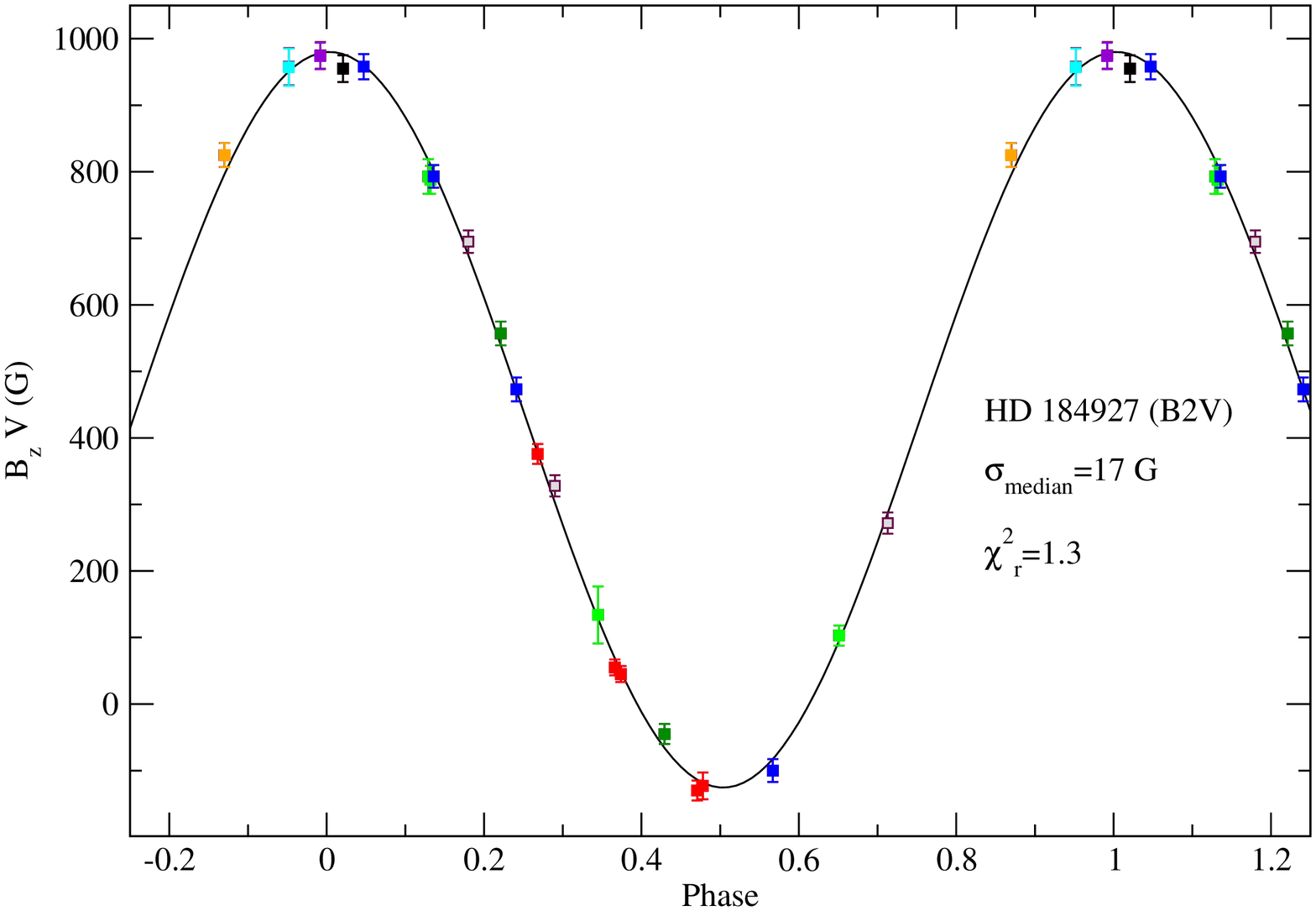}\includegraphics[width=2.2in,angle=-90]{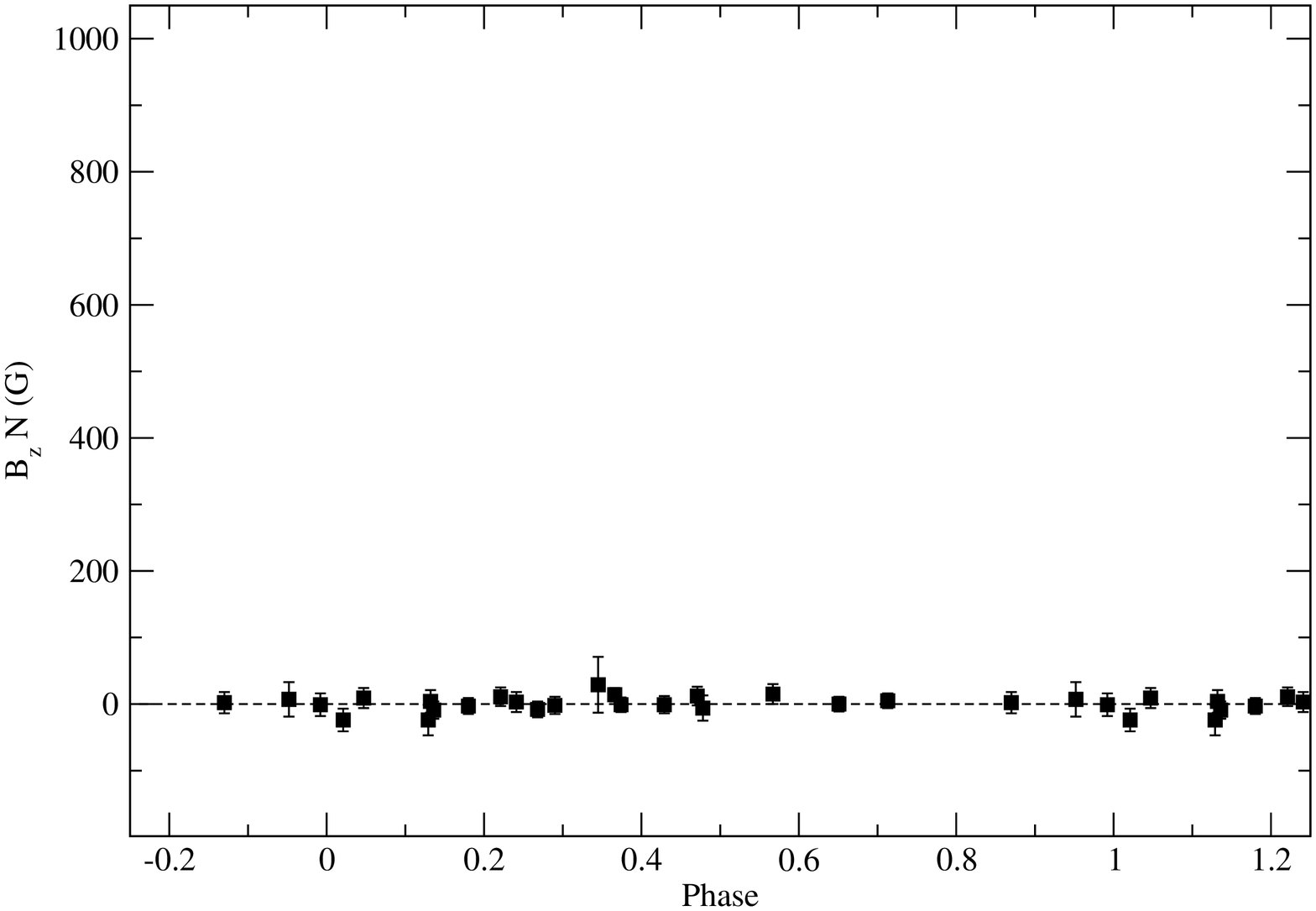}
\caption{{\em Left panel -}\ Longitudinal magnetic field variation of HD 184927, measured from Stokes $V$. Colours indicate different epochs of observation. {\em Right panel -}\ Longitudinal field of HD 184927 measured from diagnostic null profiles. The measurements are individually and statistically consistent with the absence of a magnetic field, and the low reduced $\chi^2$ supports the accuracy of the error bars.}
\end{centering}
\end{figure}

One of the major undertakings of the MiMeS Project is the detailed analysis of previously known and newly-detected magnetic massive stars as part of the TC. Datasets obtained for these stars allow us to constantly assess the reliability of our measurements, which is key to ensuring that the results of the SC are trustworthy.

MiMeS quality control focuses on the reproducibility of our measurements and verification of their associated uncertainties (both the Zeeman signatures and corresponding \bz\ measurements). This is accomplished in 3 basic ways. First, we confirm that the Stokes $V$ profile and longitudinal field are reproduced in observations of a target acquired at similar rotational phases. In Fig. 2 (left panel) we display the phased \bz\ variation for one of the MiMeS TC targets, HD 184927 (Yakunin et al., in prep), which illustrates the repeatability of our measurements over many different epochs, for a star with a rotational period of $\sim 9.5$~days \citep[e.g.][]{1979PASP...91..789L}, from observations spanning 3 years. Secondly, we employ the diagnostic null (or $N$) spectrum - to test the instrumental systems for spurious contributions to the polarization. In Fig. 2 (right panel) we display the phased \bz\ variation obtained from the diagnostic null of HD 184927, demonstrating the lack of any signal in the check spectrum.

Finally, we verify the accuracy of our uncertainties by analyzing the distribution of detection significances of the longitudinal magnetic field measurements \bz/$\sigma$, from the observed SC sample with no detectable Zeeman signatures, to ensure that the distribution is consistent with a Gaussian statistics derived from the computed error bars. This procedure is performed for both Stokes $V$ and the same measurements extracted from the diagnostic null spectra.

\begin{figure}
\begin{centering}
\includegraphics[width=2.0in,angle=-90]{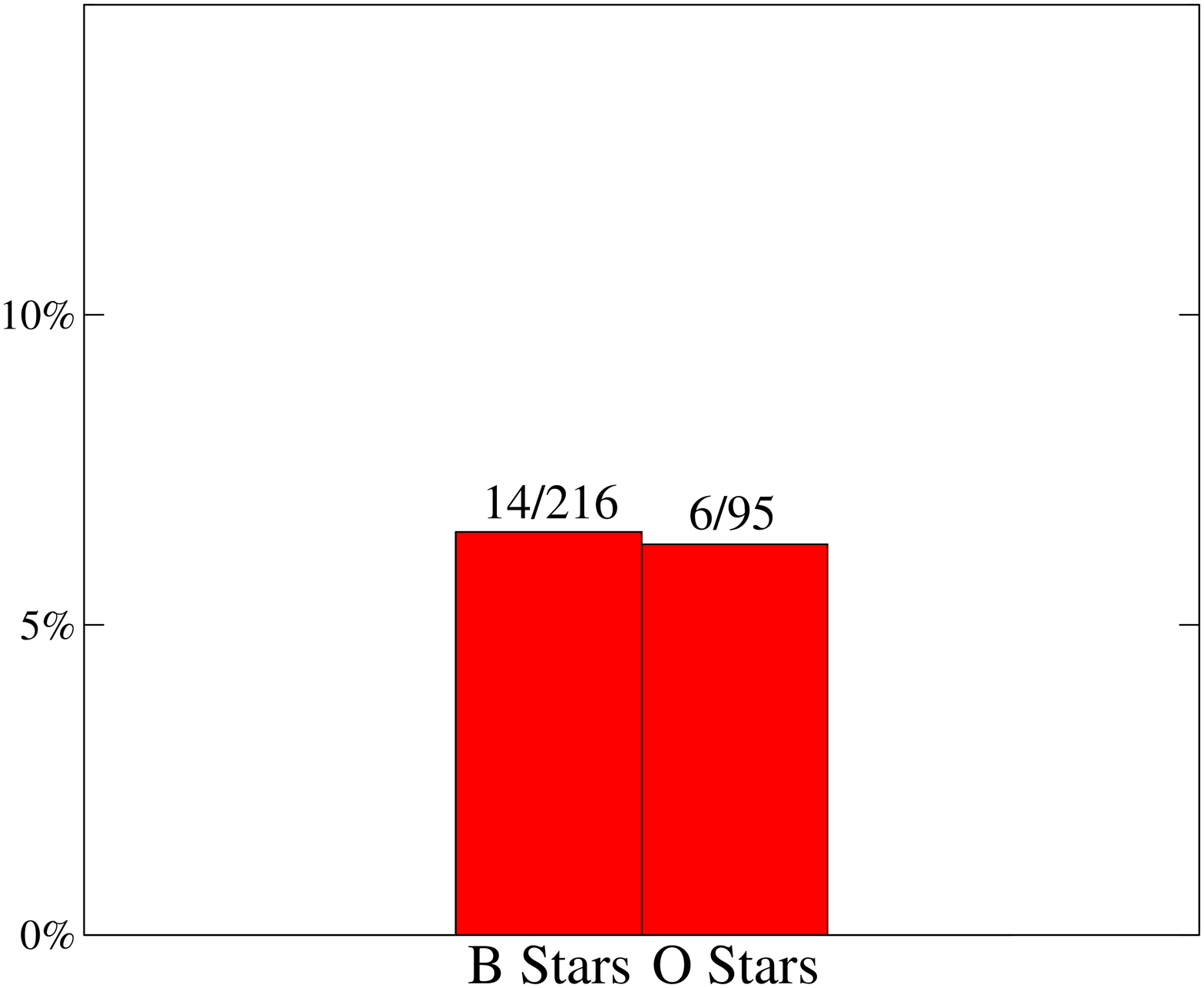}\includegraphics[width=2.0in,angle=-90]{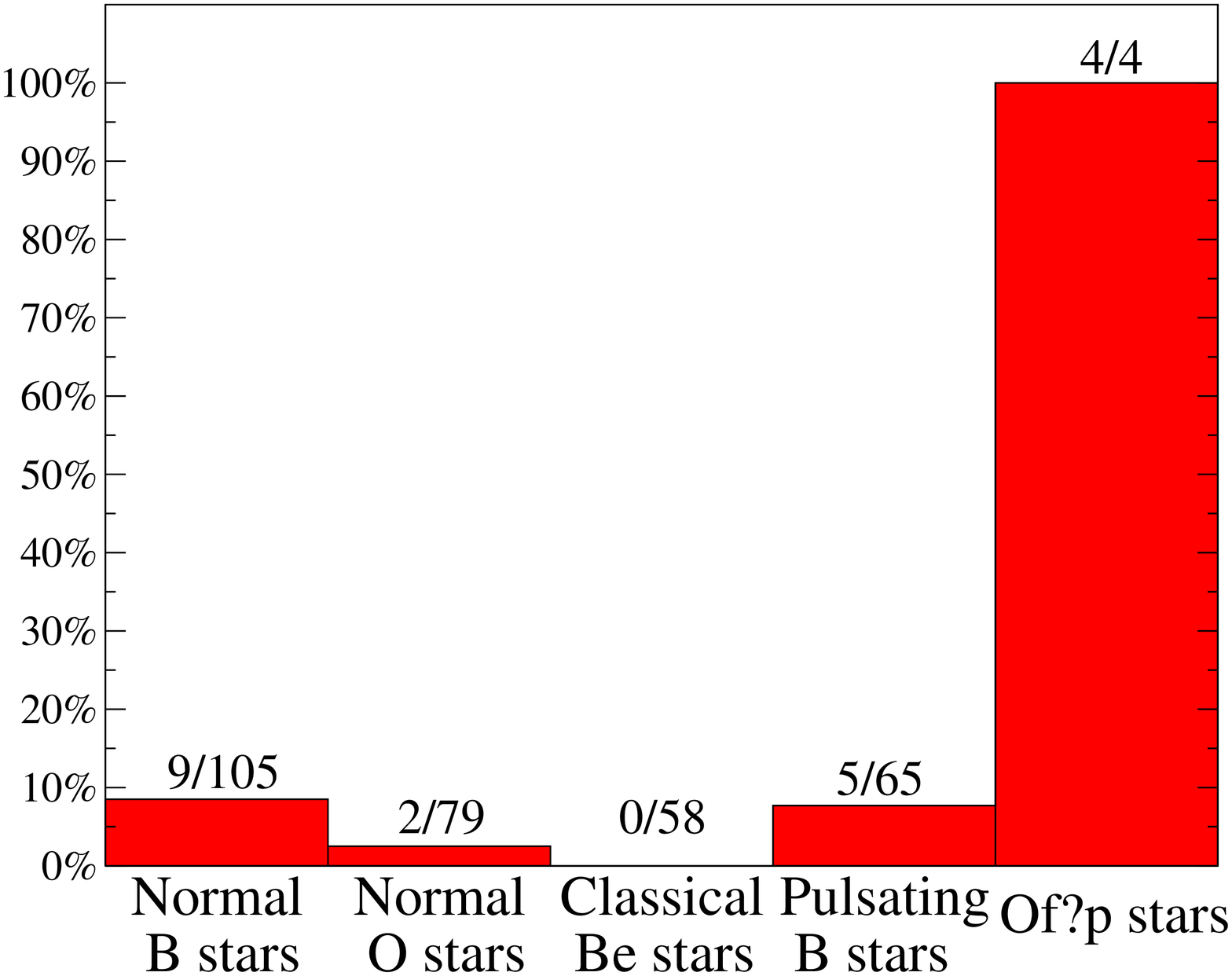}
\caption{Incidence fraction of magnetic stars relative to the total indicated sample of all stars observed as part of the MiMeS survey. Our statistics indicate a $\sim 6.5$\% incidence rate of magnetism amongst all massive stars in our sample.}
\end{centering}
\end{figure}

\section{Current results of the MiMeS survey}






%



As of March 2012, approximately 357 OB stars have been observed within the context of the MiMeS SC. This corresponds to over 1200 individual spectra (for certain targets we combine sequences of observations to improve the S/N without saturating the CCD). 

From our preliminary statistics based on 311 stars, we find that approximately 6.5\% of all O- and B-type stars observed as part of the SC and related PI programs host detectable magnetic fields. The incidence rates of magnetic fields amongst O-type versus B-type stars (left panel of Fig. 3) are essentially identical.

If we further classify the observed stars based on their spectroscopic or other fundamental properties (right panel of Fig. 3) we find that the peculiar class of massive, emission line Of?p stars stands out, with all observed stars of this class detected to be magnetic. These are small number statistics but if we include another Of?p star, which was previously found to be magnetic \citep[HD 191612;][]{2006MNRAS.365L...6D}, and is being observed as part of the TC \citep{2011MNRAS.416.3160W}, we find that all known Galactic Of?p stars are magnetic. 

We also point out the class of pulsating B-type stars, consisting of the Slowly Pulsating B-type (SPB) stars and the $\beta$~Cep pulsators. These stars are of particular interest since it has been claimed that upwards of 40\% of these stars are magnetic from low-resolution observations \citep[e.g.][]{2006MNRAS.369L..61H,2009AN....330..317H}. Our measurements are in conflict with these results as we find detectable magnetic fields only in about 7.5\% of the observed stars, which is similar to the incidence fraction ($\sim 6.5$\%) that we obtain in normal B-type stars. For a few of these stars we have obtained multiple observations and have carried out a more detailed analysis to establish the robustness of this result \citep{2012ApJ...750....2S}.

A final remarkable result is the lack of detected fields in any classical Be star. We follow the definition of classical Be stars as B-type stars close to the main sequence that exhibit line emission over the photospheric spectrum, resulting from circumstellar gas that is confined to an equatorial decretion disk that is rotating in a Keplerian orbit \citep{2003PASP..115.1153P}. This distinction allows us to investigate the potential role of magnetic fields in the formation of such circumstellar disks.
We have verified that the lack of detections in classical Be stars is not a result of systematically achieving a lower magnetic sensitivity in these stars compared to the normal B-type stars. We do find, on average, that the classical Be stars have higher rotational velocities compared to the normal B-type stars in our sample, but despite this fact, we still sample a population of classical Be stars for which we expect to have detected a small number magnetic stars, under the hypothesis that classical Be stars have a magnetic incidence fraction and magnetic field characteristics similar to the normal B-type stars. We also confirm that the lack of detections is not a result of reduced sensitivity resulting from emission contributions since we detect magnetic fields in other emission line O- and B-type stars (that do not fall into the classical Oe or Be star classification, e.g. Of?p stars and Herbig Ae/Be stars).







\begin{figure}
\begin{centering}
\includegraphics[width=2.5in]{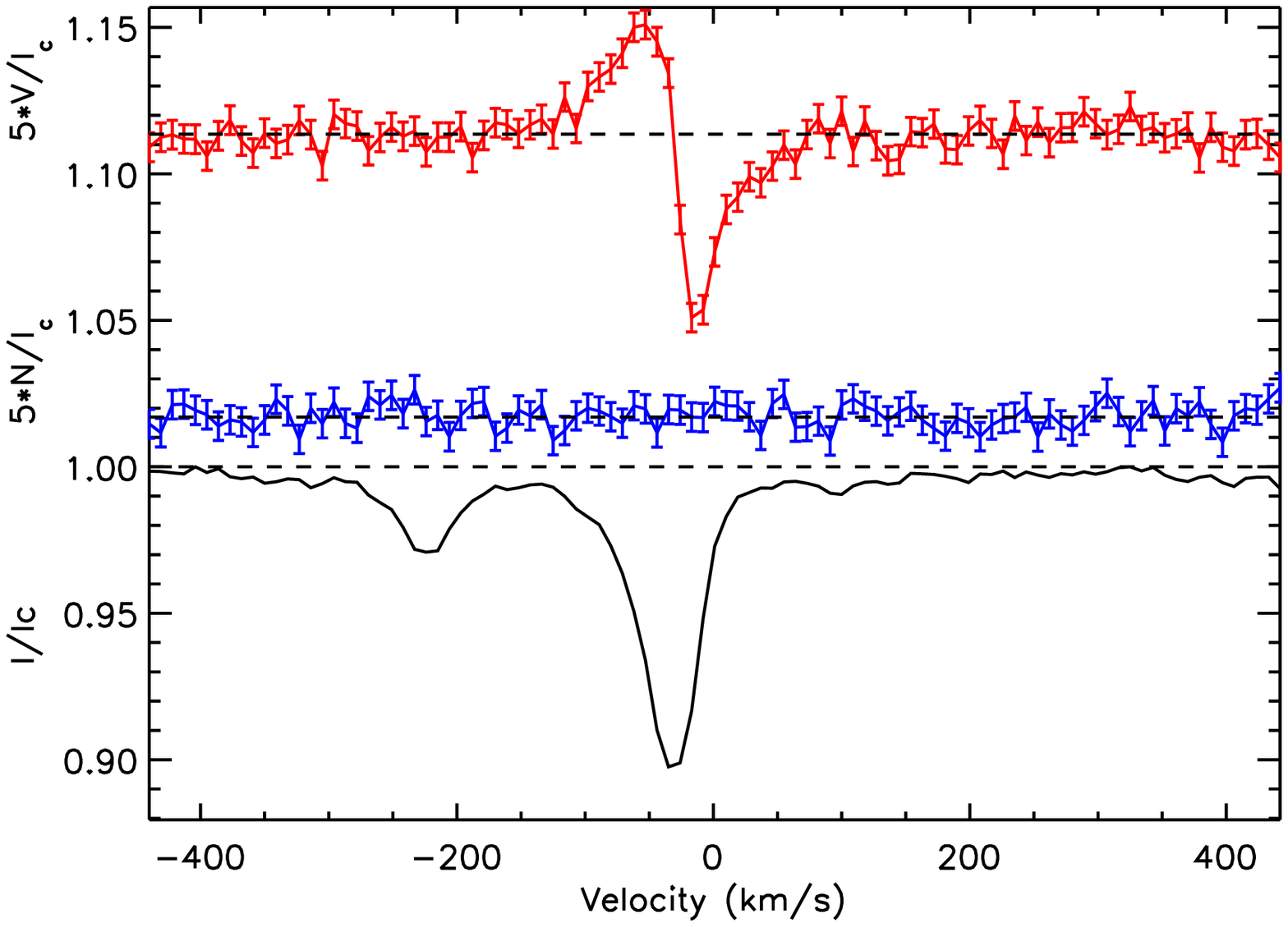}\includegraphics[width=2.5in]{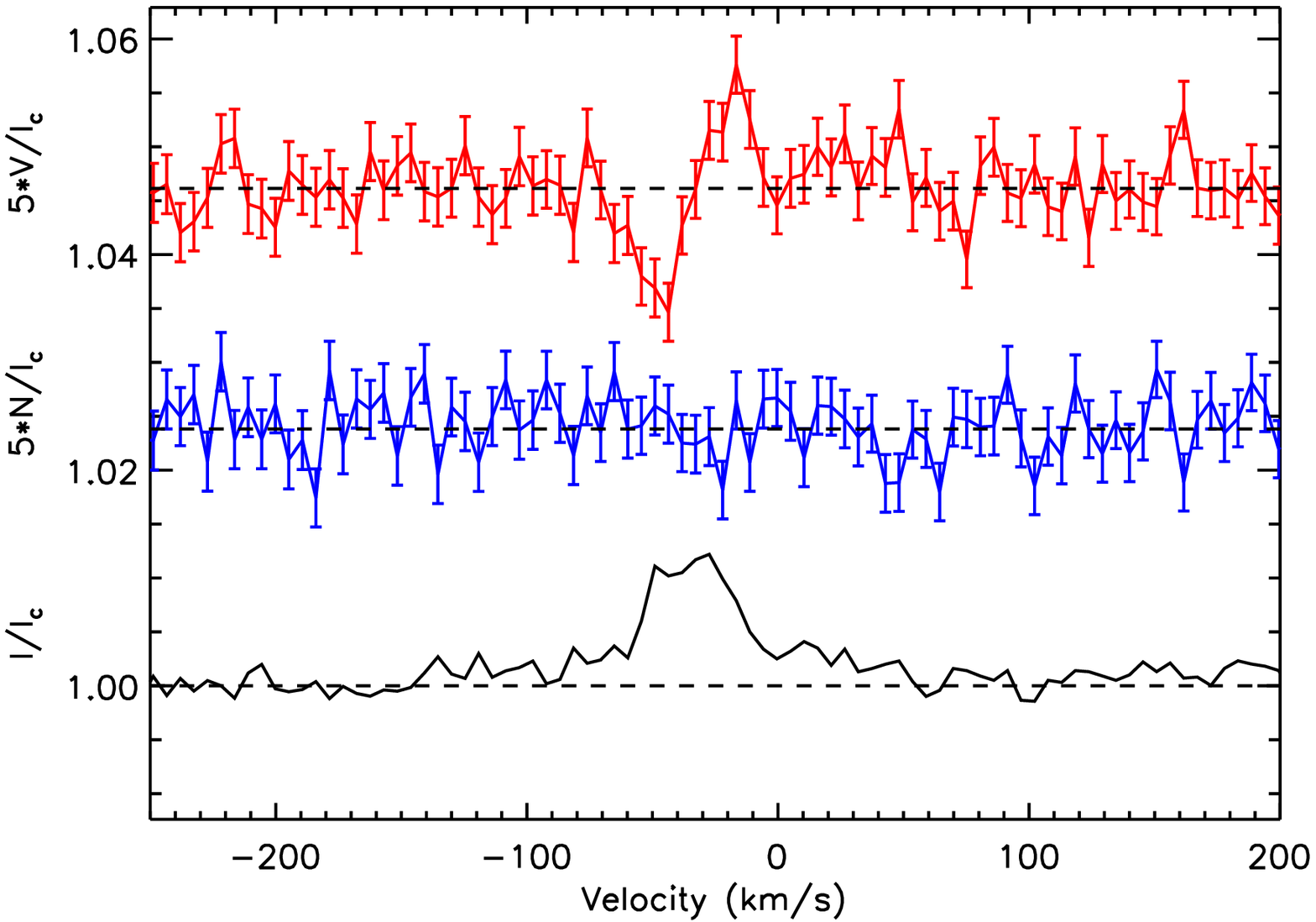}
\caption{Least-Squares Deconvolved Stokes $I$ (bottom), $V$ (top) and diagnostic $N$ profiles of NGC 1624-2. {\em Left side -}\ Profile corresponding to absorption lines of He~{\sc i} O~{\sc iii} and C~{\sc iv}. {\em Right side -}\ Profile corresponding to emission lines of O~{\sc iii}. Note the reversed Stokes $V$ profile in the right panel as compared to the left panel. From Wade et al. (2012, MNRAS, submitted).}
\end{centering}
\end{figure}

\section{Notable discoveries}

Amongst the many magnetic stars discovered or confirmed by the MiMeS survey there exist a variety of remarkable and exotic objects: extreme slow rotators \citep{2010MNRAS.407.1423M} and extreme rapid rotators \citep{2012MNRAS.419.1610G}; magnetic stars of extreme mass \citep{2012MNRAS.419.2459W}; and those with peculiar UV spectra \citep{2011arXiv1111.0982P}.

The most recent MiMeS semesters have yielded the discovery of a number of additional remarkable magnetic objects. First among these is NGC 1624-2, the faintest and most extreme of the known Galactic Of?p stars. MiMeS observations have resulted in the detection of an extraordinarily strong magnetic field (Wade et al., MNRAS, submitted). The longitudinal component of the field - over 5 kG - implies the presence of a surface dipole of approximately 20~kG - around 8 times more intense than in any other magnetic O-type star. Detection of resolved Zeeman splitting of spectral lines in Stokes $I$ spectra confirms the presence of a very intense surface magnetic field. Equally surprising is the discovery of {\em reversed} Stokes $V$ signatures in weak emission lines of light elements, in particular O~{\sc iii}. LSD profiles illustrating these properties are shown in Fig. 4. This slowly-rotating ($P_{\rm rot}=158$~d) magnetic O star will be an important target for follow-up observations during the coming months.

A second exciting result is the detection of a magnetic field in the very massive ($>100~M_\odot$) O-type binary HD 47129 (Plaskett's star; Grunhut et al., in prep.). The detection appears to be associated with the broad-lined secondary of this SB2 with $P_{\rm orb}\simeq 14.4$~d. The rapid rotation of this star places it in a unique location amongst O stars on the rotation-wind confinement diagram \citep[][Petit et al., in prep.]{2011arXiv1111.0982P}.

Finally, we have obtained a definite detection of the magnetic field of the Of?p CPD -28\,2561, recently reported by \citet{2012IBVS.6019....1H} to be magnetic based on a small number of $\sim 3\sigma$ detections of the longitudinal magnetic field. Our observations demonstrate that the longitudinal field changes sign during the star's rotation, consistent with the spectroscopic monitoring of Barb\'a et al. (in preparation).

\acknowledgements GAW and JHG acknowledge support from the Natural Science and Engineering Research Council of Canada. The MiMeS surveys are based on data acquired through generous allocations of telescope time at the Canada-France-Hawaii Telescope, the T\'elescope Barnard Lyot and the ESO 3.6m telescope.

\bibliography{wade}

\begin{thebibliography}{}
\expandafter\ifx\csname natexlab\endcsname\relax\def\natexlab#1{#1}\fi
\expandafter\ifx\csname url\endcsname\relax
  \def\url#1{\texttt{#1}}\fi
\expandafter\ifx\csname urlprefix\endcsname\relax\def\urlprefix{URL }\fi
\providecommand{\eprint}[2][]{\url{#2}}

\bibitem[{{Donati} et~al.(2006){Donati}, {Howarth}, {Bouret}, {Petit},
  {Catala}, \& {Landstreet}}]{2006MNRAS.365L...6D}
{Donati}, J.-F., {Howarth}, I.~D., {Bouret}, J.-C., {Petit}, P., {Catala}, C.,
  \& {Landstreet}, J. 2006, MNRAS, 365, L6. \eprint{arXiv:astro-ph/0510395}

\bibitem[{{Donati} et~al.(1997){Donati}, {Semel}, {Carter}, {Rees}, \& {Collier
  Cameron}}]{1997MNRAS.291..658D}
{Donati}, J.-F., {Semel}, M., {Carter}, B.~D., {Rees}, D.~E., \& {Collier
  Cameron}, A. 1997, \mnras, 291, 658

\bibitem[{{Grunhut} et~al.(2012){Grunhut}, {Rivinius}, {Wade}, {Townsend},
  {Marcolino}, {Bohlender}, {Szeifert}, {Petit}, {Matthews}, {Rowe}, {Moffat},
  {Kallinger}, {Kuschnig}, {Guenther}, {Rucinski}, {Sasselov}, \&
  {Weiss}}]{2012MNRAS.419.1610G}
{Grunhut}, J.~H., {Rivinius}, T., {Wade}, G.~A., {Townsend}, R.~H.~D.,
  {Marcolino}, W.~L.~F., {Bohlender}, D.~A., {Szeifert}, T., {Petit}, V.,
  {Matthews}, J.~M., {Rowe}, J.~F., {Moffat}, A.~F.~J., {Kallinger}, T.,
  {Kuschnig}, R., {Guenther}, D.~B., {Rucinski}, S.~M., {Sasselov}, D., \&
  {Weiss}, W.~W. 2012, \mnras, 419, 1610. \eprint{1109.3157}

\bibitem[{{Howarth} et~al.(2007){Howarth}, {Walborn}, {Lennon}, {Puls},
  {Naz{\'e}}, {Annuk}, {Antokhin}, \& {Bohlender}}]{2007MNRAS.381..433H}
{Howarth}, I.~D., {Walborn}, N.~R., {Lennon}, D.~J., {Puls}, J., {Naz{\'e}},
  Y., {Annuk}, K., {Antokhin}, I., \& {Bohlender}, D. e.~a. 2007, \mnras, 381,
  433. \eprint{0707.0594}

\bibitem[{{Hubrig} et~al.(2009){Hubrig}, {Briquet}, {De Cat}, {Sch{\"o}ller},
  {Morel}, \& {Ilyin}}]{2009AN....330..317H}
{Hubrig}, S., {Briquet}, M., {De Cat}, P., {Sch{\"o}ller}, M., {Morel}, T., \&
  {Ilyin}, I. 2009, Astronomische Nachrichten, 330, 317. \eprint{0902.1314}

\bibitem[{{Hubrig} et~al.(2006){Hubrig}, {Briquet}, {Sch{\"o}ller}, {De Cat},
  {Mathys}, \& {Aerts}}]{2006MNRAS.369L..61H}
{Hubrig}, S., {Briquet}, M., {Sch{\"o}ller}, M., {De Cat}, P., {Mathys}, G., \&
  {Aerts}, C. 2006, \mnras, 369, L61. \eprint{arXiv:astro-ph/0604283}

\bibitem[{{Hubrig} et~al.(2012){Hubrig}, {Kholtygin}, {Scholler}, {Langer},
  {Ilyin}, \& {Oskinova}}]{2012IBVS.6019....1H}
{Hubrig}, S., {Kholtygin}, A., {Scholler}, M., {Langer}, N., {Ilyin}, I., \&
  {Oskinova}, L. 2012, Information Bulletin on Variable Stars, 6019, 1

\bibitem[{{Levato} \& {Malaroda}(1979)}]{1979PASP...91..789L}
{Levato}, H., \& {Malaroda}, S. 1979, \pasp, 91, 789

\bibitem[{{Maeder} \& {Meynet}(2003)}]{2003A&A...411..543M}
{Maeder}, A., \& {Meynet}, G. 2003, \aap, 411, 543.
  \eprint{arXiv:astro-ph/0309672}

\bibitem[{{Maeder} \& {Meynet}(2004)}]{2004A&A...422..225M}
--- 2004, \aap, 422, 225. \eprint{arXiv:astro-ph/0404417}

\bibitem[{{Martins} et~al.(2010){Martins}, {Donati}, {Marcolino}, {Bouret},
  {Wade}, {Escolano}, {Howarth}, \& {Mimes
  Collaboration}}]{2010MNRAS.407.1423M}
{Martins}, F., {Donati}, J.-F., {Marcolino}, W.~L.~F., {Bouret}, J.-C., {Wade},
  G.~A., {Escolano}, C., {Howarth}, I.~D., \& {Mimes Collaboration} 2010,
  \mnras, 407, 1423. \eprint{1005.1854}

\bibitem[{{Petit} et~al.(2011){Petit}, {Kochukhov}, {Massa}, {Marcolino},
  {Wade}, \& {Ignace}}]{2011arXiv1111.0982P}
{Petit}, V., {Kochukhov}, O., {Massa}, D.~L., {Marcolino}, W.~L.~F., {Wade},
  G.~A., \& {Ignace}, R. 2011, ArXiv e-prints. \eprint{1111.0982}

\bibitem[{{Porter} \& {Rivinius}(2003)}]{2003PASP..115.1153P}
{Porter}, J.~M., \& {Rivinius}, T. 2003, \pasp, 115, 1153

\bibitem[{{Shultz} et~al.(2012){Shultz}, {Wade}, {Grunhut}, {Bagnulo},
  {Landstreet}, {Neiner}, {Alecian}, {Hanes}, \& {MiMeS
  Collaboration}}]{2012ApJ...750....2S}
{Shultz}, M., {Wade}, G.~A., {Grunhut}, J., {Bagnulo}, S., {Landstreet}, J.~D.,
  {Neiner}, C., {Alecian}, E., {Hanes}, D., \& {MiMeS Collaboration} 2012,
  \apj, 750, 2. \eprint{1202.2384}

\bibitem[{{Sundqvist} et~al.(2012){Sundqvist}, {ud-Doula}, {Owocki},
  {Townsend}, {Howarth}, \& {Wade}}]{2012MNRAS.tmpL.433S}
{Sundqvist}, J.~O., {ud-Doula}, A., {Owocki}, S.~P., {Townsend}, R.~H.~D.,
  {Howarth}, I.~D., \& {Wade}, G.~A. 2012, \mnras, L433. \eprint{1203.1050}

\bibitem[{{ud-Doula} \& {Owocki}(2002)}]{2002ApJ...576..413U}
{ud-Doula}, A., \& {Owocki}, S.~P. 2002, \apj, 576, 413

\bibitem[{{ud-Doula} et~al.(2009){ud-Doula}, {Owocki}, \&
  {Townsend}}]{2009MNRAS.392.1022U}
{ud-Doula}, A., {Owocki}, S.~P., \& {Townsend}, R.~H.~D. 2009, \mnras, 392,
  1022. \eprint{0810.4247}

\bibitem[{{Wade} et~al.(2012){Wade}, {Grunhut}, {Gr{\"a}fener}, {Howarth},
  {Martins}, {Petit}, {Vink}, {Bagnulo}, {Folsom}, {Naz{\'e}}, {Walborn},
  {Townsend}, \& {Evans}}]{2012MNRAS.419.2459W}
{Wade}, G.~A., {Grunhut}, J., {Gr{\"a}fener}, G., {Howarth}, I.~D., {Martins},
  F., {Petit}, V., {Vink}, J.~S., {Bagnulo}, S., {Folsom}, C.~P., {Naz{\'e}},
  Y., {Walborn}, N.~R., {Townsend}, R.~H.~D., \& {Evans}, C.~J. 2012, \mnras,
  419, 2459. \eprint{1108.4847}

\bibitem[{{Wade} et~al.(2011){Wade}, {Howarth}, {Townsend}, {Grunhut},
  {Shultz}, {Bouret}, {Fullerton}, {Marcolino}, {Martins}, {Naz{\'e}}, {Ud
  Doula}, {Walborn}, \& {Donati}}]{2011MNRAS.416.3160W}
{Wade}, G.~A., {Howarth}, I.~D., {Townsend}, R.~H.~D., {Grunhut}, J.~H.,
  {Shultz}, M., {Bouret}, J.-C., {Fullerton}, A., {Marcolino}, W., {Martins},
  F., {Naz{\'e}}, Y., {Ud Doula}, A., {Walborn}, N.~R., \& {Donati}, J.-F.
  2011, \mnras, 416, 3160. \eprint{1106.3008}

\end{thebibliography}

\end{document}